\documentstyle[12pt]{article}

\begin{document}

\topmargin -12mm
\oddsidemargin 5mm

\setcounter{page}{1}
\vspace{2cm}
\begin{center}

{\bf THE SEARCH OF LARGE CROSS SECTION ASYMMETRIES IN THE PAIR
PRODUCTION PROCESS BY POLARIZED PHOTONS\footnote {The work is
supported by ISTC grant A-099}}\\
\vspace{3mm}
{\large  $R.O.Avakian^{a}$,
$K.R. Dallakyan^{b}$ and $S.M. Darbinyan^{c}$}\\
\vspace{3mm}
{\em  Yerevan Physics Institute, Yerevan, 375036, Armenia}\\
{\em $^{a}E-mail: ravakian@hermes.desy.de$}\\
{\em $^{b}E-mail: kolja@yerphi.am$}\\
{\em $^{c}E-mail: simon@inix.yerphi.am$}\\
\end{center}

\vspace{3mm}
\begin{abstract}
The regions of pair particle energies and emergence angles in the
process of photoproduction of $e^{+}e^{-}$ - pairs in the coulomb
field by polarized photons are found in which a value of cross
section asymmetry is essentially large. The distributions of pair
particles on energy and on angles are calculated by the
Monte-Carlo method. The obtained results for pair particle yield
and asymmetry permit us to conclude that on their basis can be
developed a new effective method for polarized photon
polarimetry.
\end{abstract}

\indent
Recently the interest towards the methods of measuring of photon
beam linear polarization degree at energies of 10 GeV and higher
are renewed. This is due the fact that in the large centers in
USA (JLAB, photon energy 4-10 GeV) and in Europe (CERN, photon
energy 100-150 GeV) the experiments on investigation of
photoproduction processes are planned to be carried out by the
polarized photon beams [1,2]. For the monitoring of the photon
beam polarization the measurement of photon beam polarization
degree will be carried out parallel to basic experiment. Hence in
the preparation period of the experiment the suitable method for
the polarization measurement will be chosen. The following
methods for measurement of photon beam linear polarization are
well known: a) the measurement of the asymmetry $e^{+}e^{-}$ -
pair particle yield in amorphous and crystal media [3,4], b) the
measurement of the asymmetry in the angular distribution of
recoil electrons in the process of $e^{+}e^{-}$ - pair production
in the field of atomic electrons (triplet production) [5], c)
measurement of the asymmetry in nuclear reactions of deuteron
photodisntegration, the photoproduction $\pi$- mesons and the
measurement of asymmetry in angular distributions of
$\pi^{+}\pi^{-}$- mesons of the decay products photoproduced
$\rho^{0}$-mesons [6]. As polarization measurement will be
carried out in parallel with the basic experiment a concrete
requirements should be provided for the polarization measurement
method:

\indent
a) The value of the asymmetry should be sufficently large in
order to diminish influence of systematic errors.\\
\indent
b) The polarimeter should not have essential influence on the
photon beam intensity (less then $1\%$).\\
\indent
c) The method should provide small errors ($2-5\%$).\\
\indent
d) The polarimeter should be sufficiently simple, inexpensive and
reliable and should work in a full-automatic regime.\\
\indent
The comparison of the above mentioned methods brought us to the
conclusion that for the application in the JLAB experiment the
most suitable methods are the ones based on the measurement of
asymmetry in the processes of triplet photoproduction and
$e^{+}e^{-}$ - pair production in the amorphous and crystal
media. The first process is more suitable from the point of view
of the measurement methodic - one needs to measure the recoil
electron asymmetry at large angles (larger then $15\%$), however
the value of the asymmetry is not large (about $10\%$) and yield
of reaction is relatively small (proportional to Z). The value of
the asymmetry of the pair production in crystals grows with
photon energy and is essentially large at the energies
$\omega>10$ GeV. Although this method has some technical
difficulties connected with a use of goniometrical system,
magnetic spectrometer etc., nevertheless it has an essential
advantage - the value of the asymmetry does not depend on multiple
scattering of pair particles.
It should be pointed out that this method is the best
at high photon energies of order 10 GeV or more.\\
\indent
The method of measurement of asymmetry in the process of
$e^{+}e^{-}$ - pair production in amorphous media is relatively
simple [3,7]. In the experiment the asymmetry in the pair
particle yield in the certain regions of polar and azimuth angles
is measured. In review report in [1] (p.67) the known results of
this method were discussed in details. In particular the results
obtained in [7] are discussed. In this paper the asymmetry of
production of nearly coplanar pairs with respect of polarization
plane was considered. The results for the asymmetry were obtained
in the case when one pair paricle emerges in the polarization
plane or in the plane perpendicular to it and over azimuth angle
of second particle the integration is carried out in the small
interval $2\Delta\phi$ or outside of this interval (so called
wedge method). In both cases the asymmetry was calculated for
total pair yield and for symmetric pairs afther integration cross
section over polar angles. The maximal value of the asymmetry
makes up $18\%$. These investigations have found their promotion
in paper [8]. In this work the dependence of the asymmetry upon
the angle between polarization and pair emergence planes was
considered. Authors confirm that in this way the large values for
the asymmetry can be obtained ($29\%$ for $\omega=0.5$ GeV and
$24\%$ for $\omega=4$ GeV for the carbon target).\\
\indent
Here the distributions of the pair particles on energies and on
polar and azimuth angles produced by polarized photons are
calculated. The calculations are carried out for the coulomb
filed by Monte-Carlo simulation. We proceed from the formulae of
the $e^{+}e^{-}$- pair production differential cross section in
the high energy and small angle approximation [9]:
\begin{equation}
\label{AA}
d^{5}\sigma_{p}=\frac{\sigma_{0}}{\pi^{2}}\frac{y(1-y)}{\Delta^{2}}dy
(X_{unp}-\xi_{3}X_{pol})du^{2}_{+}du^{2}_{-}d\varphi_{+}d\varphi_{-},
\end{equation}
where $\sigma_{0}=Z^{2}r^{2}_{0}\alpha$, Z is the nuclear charge,
$r=e^{2}/m$ the classical electron radius $(\hbar=c=1)$,
$\alpha=1/137$, $y=\varepsilon_{\pm}/\omega$,
$\varepsilon_{\pm},\omega$ - are the positron (electron) and
photon energies, $\varphi(y)=y/(1-y)+(1-y)/y$,
$u_{\pm}=\gamma_{\pm}\vartheta_{\pm}$,
$\gamma_{\pm}=\varepsilon_{\pm}/m, \xi_{\pm}=1/(1+u^{2}_{\pm}),
\varphi_{\pm}$ are the azimuth angles, counted from the
polarization plane, $\xi_{3}$ - is the photon Stock's parameter,
\begin{equation}
\label{AB}
\Delta=\delta^{2}/m^{2}+u^{2}_{+}+u^{2}_{-}+2u_{+}u_{-}\
cos(\varphi_{+}-\varphi_{-})+\beta^{2},
\end{equation}
\begin{equation}
X_{unp}=(\xi_{+}-\xi_{-})^{2}+\frac{1}{2}\varphi(y)\xi_{+}\xi_{-}
[u^{2}_{+}+u^{2}_{-}+2u_{+}u_{-}\cos(\varphi_{+}-\varphi_{-})],
\end{equation}
\begin{equation}
\label{AD}
X_{pol}=\xi^{2}_{+}u^{2}_{+}\cos2\varphi_{+}+\xi^{2}_{-}u^{2}_{-}
\cos2\varphi_{-}+2\xi_{+}\xi_{-}u_{+}u_{-}\cos(\varphi_{+}+\varphi_{-}),
\end{equation}
$\delta=m^{2}/2\omega y(1-y)$ is the minimum momentum trasfer to
the nuclei and $(m\beta)^{-1}=R=0,4685. Z^{-1/3}({\AA})$ is the
screening radius.\\
\indent
The calculations are carried out for the photons with an energy
of 4 Gev
and for copper terget. For the verification of obtained results
we repeated the calculations of the wedge method [7] by the
Monte-Carlo simulation. The results are shown in Fig. 1 (curves a
and b). At small $\Delta\phi$ results coincide with that of [7],
but with increasing $\Delta\phi$ the discrepancy arises and
grows. Obviously this is connected with the fact, that results in
[7] are obtained in the approximation $\Delta\phi\ll 1$
neglecting terms of higher order of $\Delta\phi$ in (1). Our
results are obtained directly from (1) without additional
assumptions.\\
\indent
In [7] the integration over azimuth angle of one particle is
carried out in the interval $2\Delta\phi$ while for another
particle it is assumed $\varphi =0$ or $\varphi=\pi/2$ (i. e.
$\Delta\varphi=0$). In experiment as particles are registered in
certain intervals $\Delta\varphi$ we have investigated the
influence of $\Delta\varphi$ on results. It occurs that at
$\Delta\phi\ge\Delta\varphi$ the influence of $\Delta\varphi$
is negligibly small and becomes essential at
$\Delta\phi<\Delta\varphi$ (curvess $a^{\prime}$  and
$b^{\prime}$ in Fig. 1).\\
\indent
It is also turned out that the dependence of asymmetry upon polar
angles is sufficently sharp. In Fig. 2 the distribution of pair
particles on azimuth angles $\varphi_{\pm}$ is shown
(independently of particle charge, i. e. distribution of
vertices) in cases of selection in different regions of polar
angles. Curves illustrate the complete picture of pair particle
distribution in the vertical plane. In the case of integrated
cross sections over pair particle energies in the region
$0\le\varepsilon_{\pm}\le\omega$ and over polar angles in the
region $0\le\gamma\vartheta_{\pm}\le28$ with $\gamma=\omega/m$
the value of asymmetry is no more than $11\%$ (cuve 1) but in
case of symmetric pairs with polar angles in the region 0.2
mrad $\le\vartheta_{\pm}\le0.4$ mrad or
$1.56\le\gamma\vartheta_{\pm}\le3.13$ (ring method) the value of
asymmetry is up to $33\%$ (curve 3). The ring
$1.56\le\gamma\vartheta_{\pm}\le3.13$ includes $9\%$ of total
numbers of events. Note that the wedge method
brought to the small increase of asymmetry i. e. $18\%$ instead
of $15\%$ in [9]. Another significant consequence of pair
particle selection in polar angles is that almost rectangular
spectral distribution transforms into the Gaussian with a center
in $\varepsilon_{+}=\varepsilon_{-}=\omega/2$ (Fig. 3). The width
of Gaussian distribution decreases with narrowing polar angle
interval (with diminishing the width of ring). Curve 2 in Fig. 2
corresponds to this case, the asymmetry is about $24\%$, width of
ring is $1.56\le\gamma\vartheta_{\pm}\le3.13$. Transformation of
spectral distribution into the Gaussian is an essential effect
and brings to the important conclusion. It is well known that the
asymmetry is higher in case of symmetrical pairs ($18\%$ in [7]
and $24\%$ in [8]). But for registration of symmetrical pairs the
unwieldy setup of pair-spectrometer will be used. Practically one
needn't measure the pair particle energy if we select them in
certain polar angle regions. In Fig. 3 the pair particle spectral
distribution is shown after the integration over azimuthal
angles, polar angles are in the interval
$1.56\le\gamma\vartheta_{\pm}\le3.13$.\\
\indent
It follows from aforesaid that especially good results for the
asymmetry can be obtained combining two methods by selecting
particles in azimuth angles (the wedge method) and in polar angles
(the ring method).\\
\indent
The results of such selection are illustrated in Fig. 4 and 5.
The curves a and b in Fig. 4 are the same as curves a and b in
Fig. 1, but now paricles are selected also in interval
$1.56\le\gamma\vartheta_{\pm}\le3.13$ of polar angles. The
asymmetry is large - $44\%$. This results are obtained in the
ideal case when one particle has an azimuth angle $\varphi=0$
or $\pi/2$ (i. e. $\Delta\varphi=0$) and the energies of
symmetric pair particles are accurately equal
$\varepsilon_{+}=\varepsilon_{-}=\omega/2$. The curves showing in
Fig. 5 are more real in the sense of expected results in the
experiment. The curves are calculated with the account of
$\Delta\varphi$ ($\Delta\varphi=0.05$ rad or
$\Delta\varphi/\beta=1.98$). Curve 1 is calculated in the case of
$0\le\varepsilon_{\pm}\le\omega$ and
$1.56\le\gamma\vartheta_{\pm}\le3.13$, the asymmetry is about
$27\%$. Curve 2 corresponds to the case of
$0.475\omega\le\varepsilon_{\pm}\le0.525\omega$ and
$1.56\le\gamma\vartheta_{\pm}\le3.13$. The asymmetry is large
$43\%$. In first case the asymmetry is less but the pair yield is
higher and there is no need of pair spectrometer setup. Curve 3
is calculated for symmetric pairs with polar angles in interval
$1.56\le\gamma\vartheta_{\pm}\le3.13$. The asymmetry is about
$44\%$ (cp. with the curve a in Fig. 4).\\
\indent
When it is difficult to recover complete kinematics of the
reaction $\gamma\rightarrow e^{+}e^{-}$ (especially the photon
propagation direction) it may be suuggested a new, more
appropriate, method for asymmetry measuring (so called coordinate
method). It suggests to measure the electron and positron
energies and their coordinates $x_{\pm}$
and $y_{\pm}$ in directions parallel and
perpendicular to the polarization plane. The asymmetry is defiend
as $A=(N_{\|}-N_{\bot})/(N_{\|}+N_{\bot})$ where $N_{\|}$ and
$N_{\bot}$ are numbers of pairs with $\Delta x\ge
k\Delta y$ and $\Delta y\ge
k\Delta x$, respectively, where $\Delta x=|x_{+}-x_{-}|,
\Delta y=|y_{+}-y_{-}|$ and k is constant quantity
value of which must be chosen. Results of such calculations are
collected in Table 1. The calculations are carried out for the
statistics of 1 million produced pairs in the interval
$0\le\gamma\vartheta_{\pm}\le28$. In Table 1 numbers of pairs
$N_{\|}$ and $N_{\bot}$ and values of asymmetry and statistical
errors are presented in cases of selection of pairs in certain
intervals on energy $y=e_{\pm}/\omega$ and on angle
$\alpha=\gamma d/L$, where d is distance between electron and
positron  in the vertical plane
and L is distance between converter and detector.
Results in Table 1 show that the cross section asymmetry
increases essentially with decreasing of selection angle
$\alpha$. At a same time the statistical errors are almost the
same. As for quantity k his influence on the asymmetry is small.
Large values for asymmetry can be obtained only at $k\ge 100$.
But in this case the cross sections are negligiblity small. The
situation is same as in the wedge method [7] when over azimuth
angle $\varphi_{-}$ is carried out integration in small interval
$(-\Delta\phi, \Delta\phi)$ with $\Delta\phi/\beta<0.4$ (see
Fig.1, curve a). The coordinate method is close in essence to one
of selection of pairs on angle $\omega$ [8]. Unfortunately in [8]
it does not noted concrete conditions under which the quantities
$\sigma_{\|}$ and $\sigma_{\bot}$ are calculated. As a
consequence it is not clear that the value of asymmetry of $24\%$
is a result of transition to the angle of $\omega$ or it is
conditioned by detector size and magnitude of parameter d i.e. is
a result of selection on the angle $\alpha$.\\
\indent
The results obtained in this work can be presented as a
theoretical basis for elaboration of setup for polarized photon
polarimetry at photon energies of a few GeV. In this article some
different methods are considered for cross section asymmetry
measurement and every one have advantages and lacks in comparison
with each other. But the polarimeter can be constructed such a
way that it becomes possible to obtain value of asymmetry by this
methods simultaneously. This permits to use possibilities of each
method and decrease statistical errors.
 In the following paper we will give description of
specific setup for polarized photon polarimetry for JLAB taking
into account of experimental conditions.

\newpage
\centerline{{\bf{Figure Captions}}}

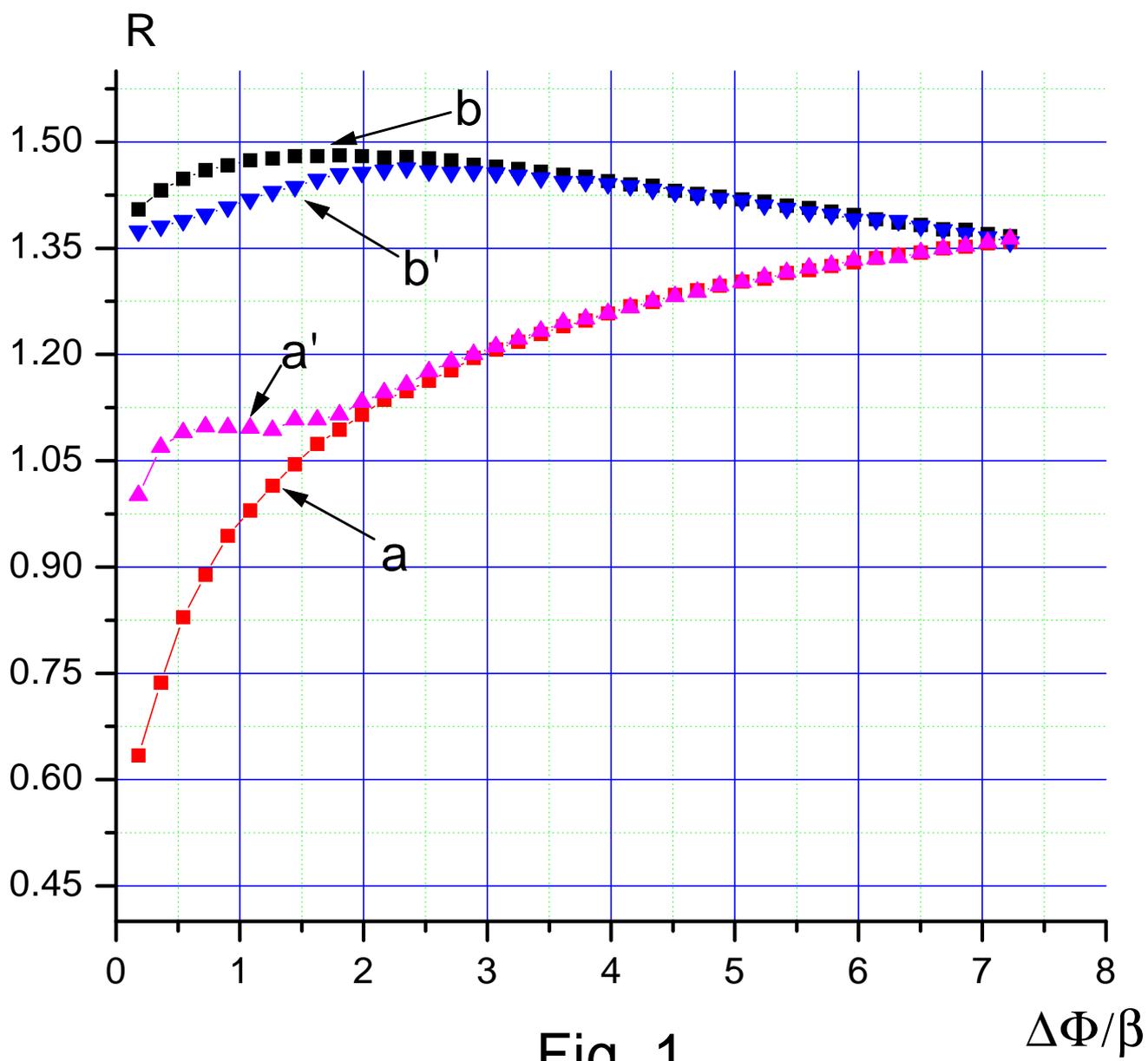
Fig.1. Dependence of the ratio
$d\sigma(\varphi_{+}=0)/d\sigma(\varphi_{+}=\pi/2)$ upon angle
$\Delta \phi/\beta$ for pairs of equal energy
$\varepsilon_{+}=\varepsilon_{-}=\omega/2$ and polar angles in
the interval $0\le\gamma\vartheta_{\pm}\le28$. Over azimuth angle
$\varphi_{-}$ the integration is carried out in the interval
($-\Delta\phi$, $\Delta\phi$) or outside this interval
($\pi-\Delta\phi$, $\pi+\Delta\phi$) (curves a and b,
correspondingly). The azimuth angles are counted out from the
polarization plane. Curves $a^{\prime}$ and $b^{\prime}$ show the same
dependencies but now the azimuth angle of first particle is in
interval $-\Delta\varphi\le\varphi_{+}\le\Delta\varphi$ or
$\pi/2-\Delta\varphi\le\varphi_{+}\le\pi/2+\Delta\varphi$ with
$\Delta\varphi/\beta=1.98$.\\
\indent
Fig.2. The distribution of pair particles on azimuth angles
$\varphi_{\pm}$. Curve 1 for $0\le\gamma\vartheta_{\pm}\le28$ and
curve 2 for $1.56\le\gamma\vartheta_{\pm}\le3.13$ in the case of
$0\le\varepsilon_{\pm}\le\omega$. Curve 3 for the case of
$1.56\le\gamma\vartheta\le3.13$ and symmetrical pairs
$\varepsilon_{+}=\varepsilon_{-}=\omega/2$.\\
\indent
Fig.3. Pair particle spectral distribution after integration over
$\varphi_{\pm}$. The polar angles are in the interval
$1.56\le\gamma\vartheta_{\pm}\le3.13$.\\
\indent
Fig. 4. Dependence of the ratio of cross sections at
$\varphi_{+}=0$ and $\varphi_{+}=\pi/2$ upon $\Delta\phi/\beta$.
The cross sections are integrated over $\varphi_{-}$ in the wedge
interval (curve a) and outside the wedge interval (curve b).
Curves are calculated in the case of $\Delta\varphi=0$,
$\varepsilon_{+}=\varepsilon_{-}=\omega/2$ and
$1.56\le\gamma\vartheta_{\pm}\le3.13$ (cp. with curves a and b in
Fig. 1.).\\
\indent
Fig. 5. The ratio of cross-sections $d\sigma(\varphi_{+})$ as in
Fig. 4. but now -$\Delta\varphi\le\Delta\varphi$
($\Delta\varphi/\beta=1.98$). Curve 1 for
$0\le\varepsilon_{\pm}\le\omega$ and
$1.56\le\gamma\vartheta_{\pm}\le3.13$. Curve 2 for
$0.475\omega\le\varepsilon_{\pm}\le0.525\omega$ and
$1.56\le\gamma\vartheta_{\pm}\le3.13$. Curve 3 for
$\varepsilon_{+}=\varepsilon_{-}=\omega/2$ and
$1.56\le\gamma\vartheta_{\pm}\le3.13$.
\newpage
\begin{table}[h!]
\centerline{\bf{Table 1.}}
\vspace{2mm}
\begin{tabular}{|l|l|l|l|l|}
\hline
\ & $0\le y\le1$ & $0\le y\le1$ & $0.475\le y\le 0.525$ &
$0.475\le y\le0.525$\\
\ & $0\le\alpha\le55$ & $1.17\le\alpha\le3.9$ & $0\le\alpha\le55$
& $1.17\le\alpha\le 3.9$\\
\hline
\ $N_{\|}(k=5)$ & 138431 & 59224 & 7258 & 4172\\
\ $N_{\bot}(k=5)$ & 113550 & 45098 & 5264 & 2738\\
\ A & $9.87\%$ & $13.54\%$ & $15.24\%$ & $20.75$\%\\
\ $\Delta A /A$ & $\pm 2\%$ & $\pm 2.31\%$ & $\pm 5.68\%$ & $\pm
5.92\%$\\
\hline
\ $N_{\|}(k=8)$ & 87469 & 37426 & 4623 & 2681\\
\ $N_{\bot}(k=8)$ & 71322 & 28248 & 3220 & 1658\\
\ A & $10.17\%$ & $13.98\%$ & $17.89\%$ & $23.58\%$\\
\ $\Delta A /A$ & $\pm 2.48\%$ & $\pm 2.82\%$ & $\pm 6.41\%$ & $\pm
6.62\%$\\
\hline
\end{tabular}
\label{tab1}
\end{table}
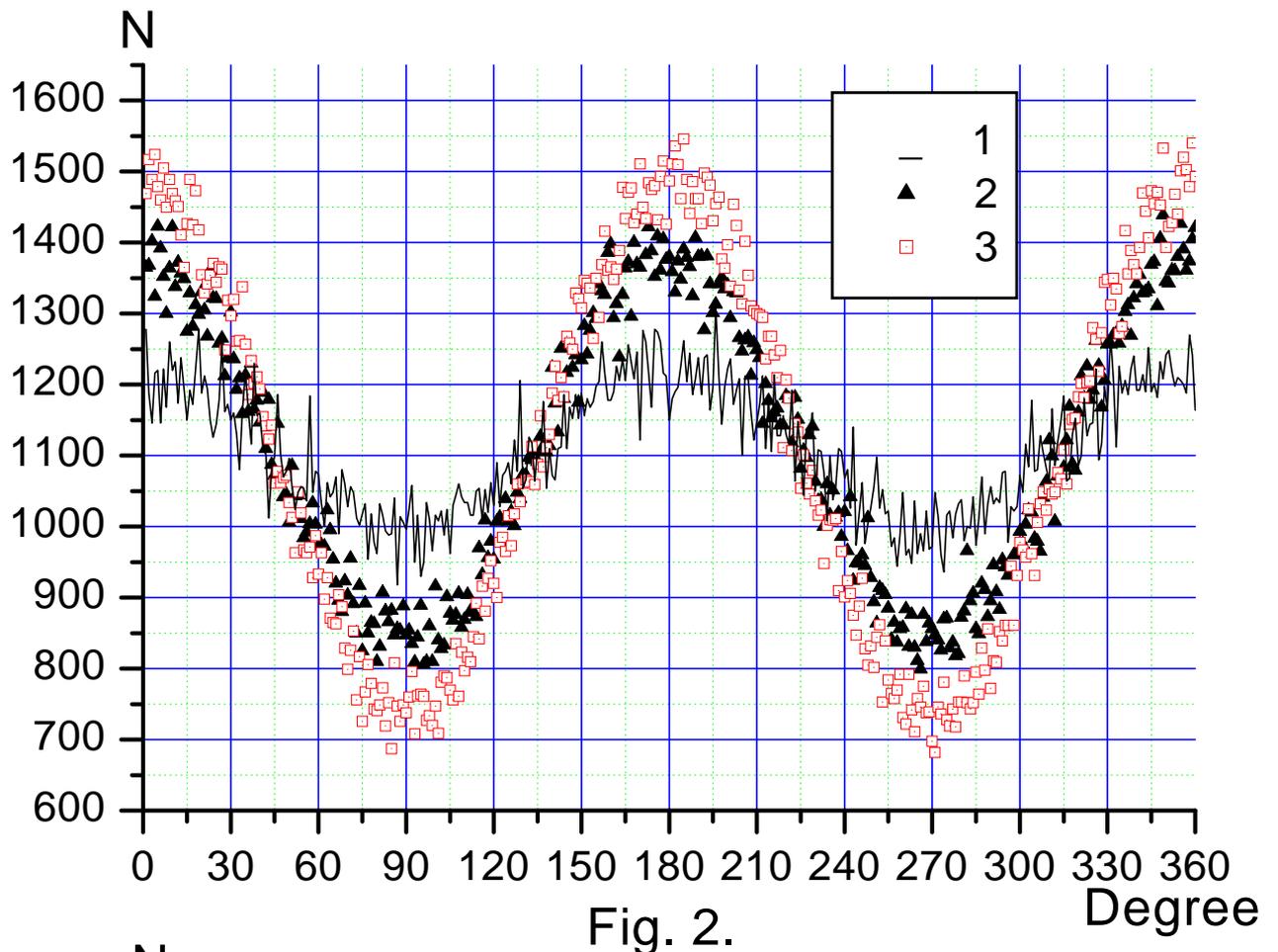
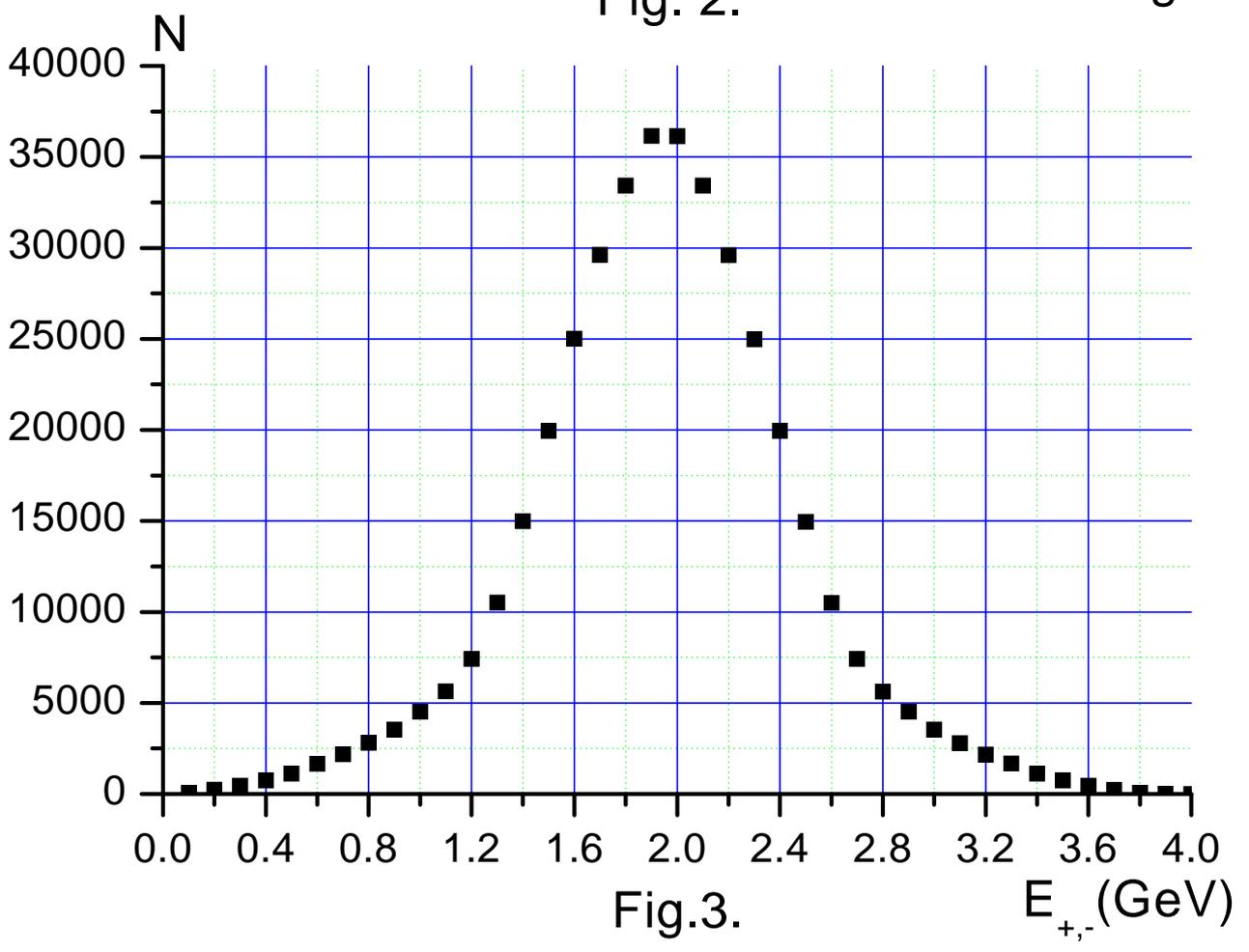
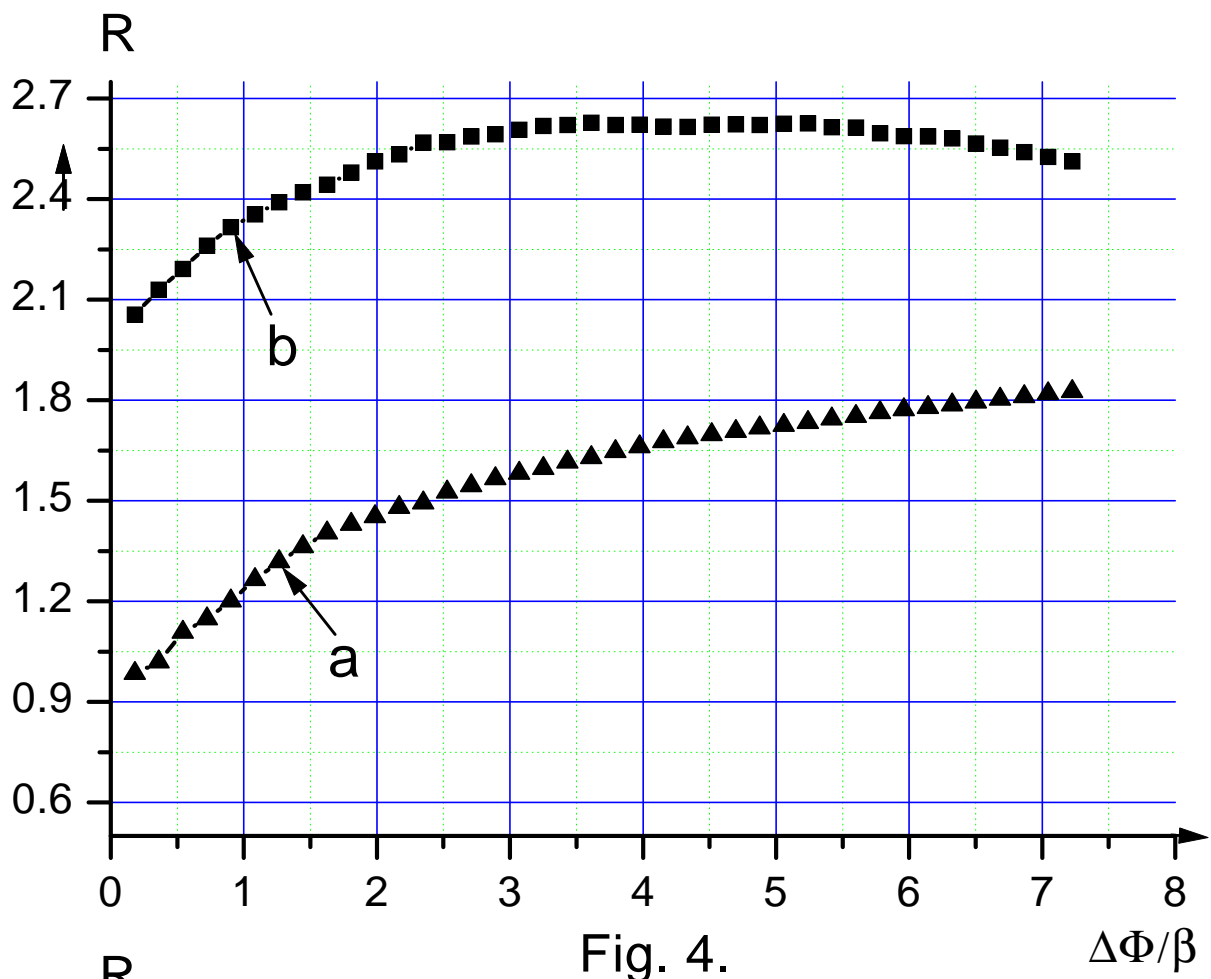
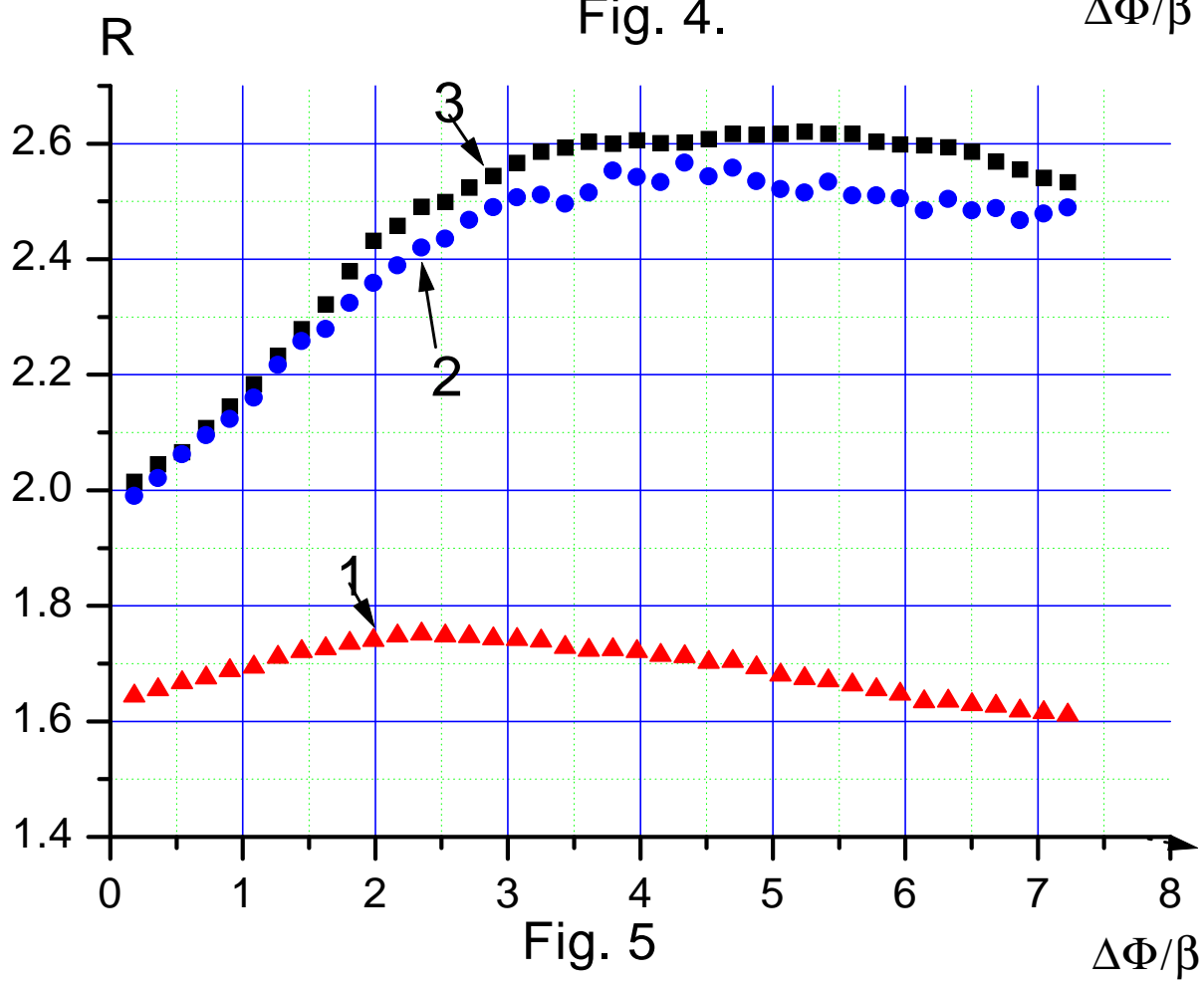
\end{document}